\documentclass{article}
\usepackage{spconf,amsmath,graphicx,hyperref, amsfonts, amssymb}

\usepackage{bm} %

\usepackage{booktabs}

\title{Low-Resource Guidance for Controllable Latent Audio Diffusion}
\ninept

\twoauthors{Zachary Novack$^{1,2\dagger}$, Zack Zukowski$^{2}$,  CJ Carr$^{2}$, Julian Parker$^{2}$, Zach Evans$^{2}$, Josiah Taylor$^{2}$,}{Taylor Berg-Kirkpatrick$^{1}$, Julian McAuley$^{1}$, Jordi Pons$^{2}$\thanks{© 2026 IEEE. Personal use of this material is permitted. Permission from IEEE must be obtained for all other uses, in any current or future media, including reprinting/republishing this material for advertising or promotional purposes, creating new collective works, for resale or redistribution to servers or lists, or reuse of any copyrighted component of this work in other works.}}
\address{$^{1}$UC -- San Diego, $^{2}$Stability AI, $^\dagger$Work done while an intern at Stability AI}

\begin{document}

\maketitle
\begin{abstract}
Generative audio requires fine-grained controllable outputs, yet most existing methods require model retraining on specific controls or inference-time controls (\textit{e.g.}, guidance) that can also be computationally demanding. By examining the bottlenecks of existing guidance-based controls, in particular their high cost-per-step due to decoder backpropagation, 
we introduce a guidance-based approach through selective TFG and Latent-Control Heads (LatCHs), which enables controlling latent audio diffusion models with low computational overhead. 
LatCHs operate directly in latent space, avoiding the expensive decoder step, and requiring minimal training resources (7M parameters and $\approx$\,4 hours of training). Experiments with Stable Audio Open demonstrate effective control over intensity, pitch, and beats (and a combination of those) while maintaining generation quality. 
Our method balances precision and audio fidelity with far lower computational costs than standard end-to-end guidance. Demo examples can be found at \href{https://zacharynovack.github.io/latch/latch.html}{\texttt{zacharynovack.github.io/latch/latch.html}}.
\end{abstract}

\section{Introduction}
\label{sec:introduction}

\begin{figure*}
    \centering
    \includegraphics[width=\linewidth]{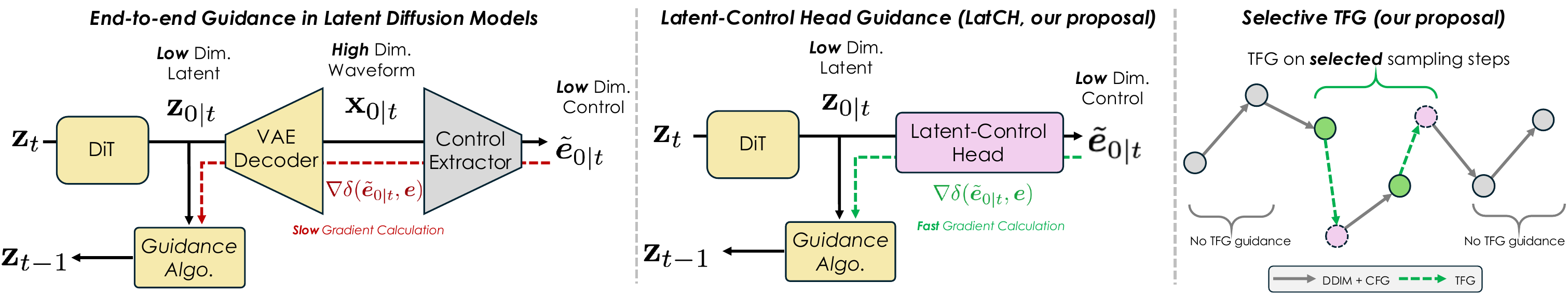}
    \caption{\textbf{\textit{Left.}} End-to-end guidance can be slow and VRAM intensive as it requires backpropagating through the VAE decoder. \textbf{\textit{Center.}} LatCH is compute efficient as it directly predicts control features from the latent space. 
    \textbf{\textit{Right.}} Selective TFG is also compute efficient as it allows applying TFG guidance only on selected sampling steps. }
    \label{fig:mainmethod}
\end{figure*}

Generative models have rapidly advanced across audio, enabling the production of coherent sonic generations from text \cite{stableaudio,evans2024open}. However, there is a growing need for controllable audio generation, as creative workflows demand fine-grained manipulations. Recent efforts rely on local-conditioning with chords, rhythms, or pitch~\cite{Wu2023MusicCM, garcia2025sketch2sound}, global-conditioning with style or reference-audio embeddings~\cite{Tal2024JointAA,rouard2024audio}, video-conditioning \cite{pascual2024masked}, editing~\cite{tsai2024audio}, or for generating stems from context (\textit{e.g.}, other stems) \cite{parker2024stemgen, nistal2024diff}.
While these enable user-steerable generation, such (conditional) models require  supervised training or finetuning with data that is challenging to collect. %

Given the high costs of training or finetuning generative models, {inference-time} control methods have been recently explored. Text-to-image research has focused on heuristic-based inference-time methods~\cite{hertz2022prompt}, optimization-based methods~\cite{novack2024ditto, novack2024ditto2}, and guidance-based methods~\cite{luo2024readoutguidance,dhariwal2021diffusion, yu2023freedom, Chung2022DiffusionPS, he2024manifold, songlgd2023, bansal2023universal, ye2024tfg}. 
We also focus on guidance-based methods, which use the gradient of a target distance function with respect to the diffusion process to guide sampling.
Training-Free Guidance~{(TFG)}~\cite{ye2024tfg} unifies most guidance-based frameworks (DPS~\cite{Chung2022DiffusionPS}, MPGD~\cite{he2024manifold}, LGD~\cite{songlgd2023}, UGD~\cite{bansal2023universal}) within a shared hyperparameter space. Yet, such approaches have only been explored for end-to-end audio-based guidance~\cite{Levy2023ControllableMP}, that is costly due to the nature of backpropagation through audio decoders during sampling. {Readouts}~\cite{luo2024readoutguidance} is a promising idea, from the image domain, that addresses this challenge by adapting guidance to be based on latent diffusion features (instead of decoded images).

We build off these ideas and show the feasibility of low-resource guidance (TFG-based) with \emph{trainable} latent-control heads (readouts-inspired).
Such \textbf{Lat}ent-\textbf{C}ontrol \textbf{H}eads (\textbf{LatCHs}) enable low-resource guidance in latent diffusion, as they operate in latent space rather than in signal space. Hence, backpropagation through the latent decoder is not required (Fig. \ref{fig:mainmethod}). Further, LatCHs are lightweight models ($\approx$\,7M parameters) that can be trained in $\approx$\,4 hours on a {single} GPU, making such models more tractable to train than fully conditional generative models. Finally, we also propose \textbf{selective TFG}, which only applies TFG guidance at few, selected diffusion steps (Fig. \ref{fig:mainmethod}). This design choice drastically improves both runtime efficiency and overall quality, as guidance becomes less prone to overoptimizing the target control.
We apply our low-resource guidance framework to Stable Audio Open (SAO) \cite{evans2024open} across three musical controls (intensity, pitch, and beats), and find that our framework successfully balances control precision with audio fidelity better than similar low-resource approaches and is more compute efficient than standard end-to-end guidance.

\section{BACKGROUND AND OUR METHOD}
\label{sec:methodology} 

We seek guidance-based methods to control latent audio diffusion models without significantly compromising quality or runtime latency. 
We do this through two methodological contributions (Fig.~\ref{fig:mainmethod}): {(1)}~{{selective TFG}} (in Sec.~\ref{sec:sTFG}) where we extend the recently proposed guidance framework TFG~\cite{ye2024tfg} with support for guidance in only few, selected diffusion steps; and {(2)} {latent-control heads} (LatCHs, in Sec.~\ref{sec:latchs}) which map the latent space of the generative model directly to the target control for order-of-magnitude latency speedups. To the best of our knowledge, we are the first to use latent-control heads for guidance in few, selected diffusion steps.

\subsection{Latent Audio Diffusion Background} 

Given text $\bm{c}$ and stereo audio $\mathbf{x}_0 \in \mathbb{R}^{2 \times TS}$ where $T$ is the length (47.55s) and $S$ is the sampling rate (44.1kHz), we wish to sample from $p(\mathbf{x}_0 | \bm{c})$. In practicality, we first embed the high-dimensional waveform into a compressed latent space using a pretrained autoencoder $\mathcal{E}$, $\mathcal{D}$ such that $\mathbf{z}_0 = \mathcal{E}(\mathbf{x}_0)$. We first define a forward process that gradually corrupts clean latents $\mathbf{z}_0$ (data to noise) according to $\mathbf{z}_t = \alpha_t\mathbf{z}_0 + \sigma_t\epsilon$, where $\epsilon \sim \mathcal{N}(0, \bm{I})$, $\alpha_t$ and $\sigma_t$ define the noise schedule, and $\mathbf{z}_1$ corresponds to fully Gaussian noise. This process then admits a natural {reverse} process (noise to data) defined by~\cite{anderson1982reverse}:
 \begin{equation}\label{eq:sde}
     \mathrm{d}\mathbf{z}_t = \left(f(\mathbf{z}_t, t) - g(t)^2 \nabla_{\mathbf{z}_t}\log p_t(\mathbf{z}_t \mid \bm{c})\right)\mathrm{d}t + g(t)\mathrm{d}\mathbf{w},
 \end{equation}
 where $\nabla_{\mathbf{z}_t}\log p_t(\mathbf{z}_t)$ is the \emph{score} of the marginal distribution of $\mathbf{z}$ at time $t$, $\mathbf{w}$ is a Weiner process, and $f$ and $g$ are the drift and diffusion coefficients defined by $\alpha_t$ and $\sigma_t$. Thus, the goal of score-based latent diffusion is to learn a model $\bm{s}_{\bm\theta}(\mathbf{z}_t, t, \bm{c}) \approx \nabla_{\mathbf{z}_t}\log p_t(\mathbf{z}_t\mid \bm{c})$ that approximates the time-dependent score and numerically integrates Eq.~\ref{eq:sde} (starting at $\mathbf{z}_1$) to achieve samples from $p(\mathbf{z}_0 \mid \bm{c})$ (and hence $\mathbf{x}_0$ through the decoder $\mathcal{D}$). In our work, we use the \emph{v-diffusion} parameterization and learn a v-prediction model $\mathbf{v}_{\bm\theta}(\mathbf{z}_t, t, \bm{c})$ such that $\bm{s}_{\bm\theta}(\mathbf{z}_t, t, \bm{c}) = - (\mathbf{z}_t + \frac{\alpha_t}{\sigma_t} \mathbf{v}_{\bm\theta}(\mathbf{z}_t, t, \bm{c}))$. Additionally, for numerical integrators we rely on the DDIM sampler~\cite{song2020denoising}:
\begin{equation}\label{eq:ddim}
    \mathbf{z}_{t-1} = \alpha_{t-1}\mathbf{z}_{0\mid t} + \sqrt{\sigma_{t-1}^2 - \eta_{t}^2} \bm{\varepsilon}_t + \eta_t\epsilon,
\end{equation}
where $\mathbf{z}_{0\mid t} = \alpha_t\mathbf{z}_{t} - \sigma_t\mathbf{v}_{\bm\theta}(\mathbf{z}_t, t, \bm{c})$ is the predicted clean data at time $t$, $\bm{\varepsilon}_t = \sigma_t\mathbf{z}_{t} + \alpha_t\mathbf{v}_{\bm\theta}(\mathbf{z}_t, t, \bm{c})$ is the predicted noise at time $t$, $\epsilon \sim \mathcal{N}(0, \bm{I})$, and $\eta_t$ are time-dependent noise parameters.

\subsection{TFG Background}
\label{sec:TFG}

TFG~\cite{ye2024tfg} aims at sampling from $p(\mathbf{x}_0 \mid \bm{c}, \bm{e})$ given a pretrained latent diffusion model $\bm{s}_{\bm\theta}(\bm{z}_t,t,\bm{c})$, where $\bm{e}$ is some additional control signal (such as RMS energy) that the diffusion model has \emph{not} been trained on. Additionally, let $\mathbf{C}(\cdot): \mathbb{R}^{2 \times TS} \rightarrow \mathbb{R}^d$ denote a differentiable $d$-dimensional feature extractor such that $\bm{\hat{e}} = \mathbf{C}(\mathbf{x}_0)$, and $\mathcal{\delta}(\bm{\hat{e}}, \bm{e})$ be some chosen distance function to evaluate how well our generated sample $\mathbf{x}_0$ follow the control signal $\bm{e}$. In evaluating the score of this probability (in the autoencoder latent space), we have:
 \begin{align}\label{eq:dps}
     \nabla_{\mathbf{z}_t}&\log p(\mathbf{z}_t \mid \bm{c}, \bm{e}) = \nabla_{\mathbf{z}_t}\log p(\mathbf{z}_t \mid \bm{c}) +  \nabla_{\mathbf{z}_t} \log p(\bm{e} \mid \mathbf{z}_t, \bm{c}) \nonumber\\
     &\approx \bm{s}_{\bm\theta}(\mathbf{z}_t, t, \bm{c}) - \rho_t\nabla_{\mathbf{z}_t}\mathcal{\delta}(\tilde{\bm{e}}_{0\mid t}, \bm{e}) - \mu_t\nabla_{\mathbf{z}_{0\mid t}}\mathcal{\delta}(\tilde{\bm{e}}_{0\mid t}, \bm{e}),
 \end{align}
where $\tilde{\bm{e}}_{0\mid t}=\mathcal{N}(\mathbf{C}(\mathcal{D}(\mathbf{z}_{0\mid t})), \gamma_t)$ is the predicted feature $\hat{\bm{e}}_{0\mid t}=\mathbf{C}(\mathbf{x}_{0\mid t})$ given our estimate of the output $\mathbf{x}_{0\mid t}=\mathcal{D}(\mathbf{z}_{0\mid t})$ at time $t$ convolved with gaussian noise $\mathcal{N}(\cdot)$ with variance $\gamma_t$. %
We decompose $\rho_t$, $\mu_t$, $\gamma_t$ as follows  $\rho$$\cdot$$s(t)$, $\mu$$\cdot$$s(t)$, $\gamma$$\cdot$$s(t)$ (respectively) where $s(t) = \alpha_t / \sum_{t=1}^T \alpha_t$ is defining their temporal structure \cite{ye2024tfg}. 
The first gradient term in Eq.~\ref{eq:dps} is known as ``variance guidance"~\cite{ye2024tfg} and is controlled by $\rho_t$, where the derivative is taken with respect to $\mathbf{z}_t$ and passes through the diffusion model. The second gradient term is known as ``mean guidance"~\cite{ye2024tfg} and is controlled by $\mu_t$, where the derivative is taken with respect to the predicted mean $\mathbf{z}_{0\mid t}$, which notably does not propagate back into the diffusion model and is thus much faster to compute.
In practice, the loss gradients in Eq.~\ref{eq:dps} can be directly subtracted from $\mathbf{z}_{t-1}$ (for ``variance guidance'') and $\mathbf{{z}}_{0|t}$ (for ``mean guidance'') after and before applying the Eq.~\ref{eq:ddim} update, thus acting as a single step of gradient descent on each optimization target.
TFG introduces two additional algorithmic modifications:
\begin{itemize}
    \item \textit{Iteration on $\mathbf{z}_{0\mid t}$} --- As ``mean guidance'' is a standard optimization step on $\mathbf{z}_{0 
    \mid t}$, it can be repeated some $N_{\mathrm{iter}}$ times each sampling step for better control alignment.
    \item \textit{Guidance recurrence} --- Similarly, the entire guidance procedure can be repeated by time-traveling \cite{yu2023freedom} back to $\mathbf{z}_t$ from $\mathbf{z}_{t-1}$ by sampling $p(\mathbf{z}_t \mid \mathbf{z}_{t-1})$ (\textit{i.e.} sampling from the forward process) for $N_{\mathrm{recur}}$ times each sampling step.
\end{itemize}
This defines the search space of TFG with $\rho, \mu, \gamma$ and $N_{\mathrm{iter}}, N_{\mathrm{recur}}$. Intuitively, $\rho$ and $\mu$ adjust the guidance strength, and $\gamma$ adjusts the accuracy strength of the guidance by adding noise to $\hat{\bm{e}}_{0\mid t}$. Finally, $N_{\mathrm{iter}}$ and $ N_{\mathrm{recur}}$ set the number of times we iteratively apply guidance.
We also experiment guiding with multiple control signals (\textit{e.g.}, jointly using RMS energy and pitch contour). To handle multiple controls, TFG averages the loss for each control, with a control dependent weight to balance different loss scales. Note that this entire framework is also compatible with Classifier-Free Guidance~(CFG)~\cite{ho2022classifier}, which we also use in our experiments.

\subsection{Selective TFG (our proposal)}
\label{sec:sTFG}
 
One key factor of guidance is omitted from TFG's design space: \emph{which} diffusion steps to apply guidance at all. Recent work shows that standard CFG \cite{ho2022classifier} can be applied only in a limited interval of diffusion steps, improving outputs and reducing computational burden~\cite{kynkaanniemi2024applying}. Combined with the fact that different guidance strengths can be used for different parts of the sampling process \cite{yu2023freedom}, this idea naturally extends to the TFG framework. Because any particular feature defined by $\mathbf{C}(\cdot)$ may perceptually arise at different points of the sampling process, limiting the guidance to only such an area reduces the risk of applying guidance in regions where it may cause the output to deviate from the data manifold. By only applying TFG guidance on selected sampling steps, we significantly reduce the computational overhead incurred by TFG while offering a clear tradeoff between control accuracy and audio quality. For example, applying TFG to more steps allows better control alignment, but increases the risk of drifting off-manifold. However, applying TFG to fewer steps reduces the risk of drifting off-manifold, but may result in a less precise control. We term this idea \textit{selective} TFG (Fig. \ref{fig:mainmethod}), where TFG is augmented with a set of binary scalars $\{\tau_i\}_{i=1}^T$ denoting which \textit{selected} steps include TFG ($T$ is the number of sampling steps).

\subsection{Latent-Control Heads (LatCHs, our proposal)}
\label{sec:latchs}

While end-to-end guidance has been explored for latent diffusion image models, its development for latent diffusion audio models~\cite{Levy2023ControllableMP, ye2024tfg} remains relatively limited—possibly due to one major practical issue: backpropagation through audio decoders $\mathcal{D}$ is \emph{exceedingly} costly. 
Note that our VAE encodes 47.55s of 44.1kHz stereo audio into a 1024-length 64-channel latent at 21.5Hz (64x downsampling rate). %
Thus, every end-to-end guidance calculation must backpropagate through $\mathcal{D}$, which is highly compute intensive in both increasing the inference latency considerably and the VRAM footprint, making it impractical (if not impossible) to run efficiently
(Fig. \ref{fig:mainmethod}).

To remedy this, we explore training latent-control heads $\mathbf{c}_{\bm\phi}(\mathbf{z}_{0})$ that avoid going through the decoder $\mathcal{D}$ to predict the control features, and instead map latents to controls directly: %
\begin{equation}\label{eq:latch}
     \mathbf{C}(\mathcal{D}(\mathbf{z}_{0})) \approx \mathbf{c}_{\bm\phi}(\mathbf{z}_{0}).
\end{equation}
By avoiding the end-to-end latent$\rightarrow$audio$\rightarrow$control mapping (Fig.~\ref{fig:mainmethod}) and directly learning to map latent$\rightarrow$control, guidance calculation can run orders of magnitude faster and cheaper as long Eq.~\ref{eq:latch} holds.

\textit{Noise-conditioned training} --- LatCHs are effectively latent discriminative models (\textit{i.e.},~an `intensity' LatCH is really just an RMS regressor on latents), yet we primarily care in how they can be used for guidance, which is defined on progressively denoised outputs. This creates a mismatch between training, where naïvely we train to map clean latents to target features, and inference, where noisy latents are used as input to the LatCHs. We explore two solutions:
\begin{itemize}
    \item \textit{Forward-Simulated Noise Conditioning (LatCH-F)}: We equip LatCHs with an additional timestep input $\mathbf{c}_{\bm\phi}(\mathbf{z}_{0|t}, t)$ through a fourier projection of $t$ and sequence concatentation. We train the noise-conditioned LatCH to predict controls given noisy latents $\mathbf{z}_t$ corrupted by the forward diffusion process. 
    \item \textit{Backwards-Simulated Noise Conditioning (LatCH-B)}: Similar to forward noise conditioning, we equip the model with timestep input. However, instead of training on forward diffused (latents + noise), we generate trajectories $\mathbf{z}_T, \mathbf{z}_{t-1}, \dots \mathbf{z}_0$ from our latent audio diffusion model and train the LatCH to map these paired intermediate steps to the features extracted from the generated output. This, in theory, exactly matches the noise distribution seen at inference by training on \emph{generated} intermediate steps rather than ones simulated from the forward diffusion process.
\end{itemize}

\section{Experiments}

We use Stable Audio Open (SAO) \cite{evans2024open} for our experiments, which  is a text-conditioned latent audio diffusion model. For our experiments, we choose the following feature extractors $\mathbf{C}(\cdot)$:
\begin{itemize}
    \item {\textit{Intensity}}: {{Root mean square}} (RMS) energy is used to model loudness  dynamics (frame length 8193, hop size 2048).
    We model intensity in decibels, and is convolved with a savitzky-golay filter~\cite{novack2024ditto} to reduce high-frequency oscillations. 
    \item \textit{Beats}: for beats alignment we use the All-In-One structure analyzer \cite{kim2023all}. Specifically, we use its beat tracking mode (which predicts time-wise binary beat probabilities).
    \item {\textit{Pitch}}: We extract monophonic pitch using CREPE \cite{kim2018crepe}, which predicts a vector of pitch probabilities per timestep.

\end{itemize}

\noindent The distance function $\mathcal{\delta}(\cdot)$ is set to binary cross entropy (BCE), except for \textit{intensity} which mean squared error (MSE).
All extractors $\mathbf{C}(\cdot)$ operate on mono audio. With that, SAO can now be guided to generate audio following specific dynamics, pitch, and beats.

\begin{table*}[h!]
    \centering
    \caption{\textbf{Qualitative and quantitative results.} $\downarrow$ the lower the better or $\uparrow$ the higher the better. }
    \resizebox{0.8\width}{!}{%
    \begin{tabular}{ll||ccc|ccc|cc||ccc}
    \toprule
         \multicolumn{2}{c||}{} & \multicolumn{3}{c|}{\textit{Quality metrics}} & \multicolumn{3}{c|}{\textit{Control alignment metrics} ($\downarrow$)} & \multicolumn{2}{c||}{\textit{Computational costs} ($\downarrow$)} & \multicolumn{3}{c}{\textit{Qualitative results} (MOS $\uparrow$)} \\ \cmidrule{3-13}
         &  & FD\textsubscript{openl3}  &  KL\textsubscript{passt} & CLAP &  Intensity&  Pitch& Beats & runtime on & VRAM & Audio & Prompt & Control\\
         \textit{Control(s)}&  \textit{Method}&  ($\downarrow$)&   ($\downarrow$)&   ($\uparrow$)&   MSE&  BCE & BCE  & GPU (s) & usage (GB) & quality & adherence & alignment\\
         \midrule
         -- & SAO \cite{evans2024open} & 96.51 & 0.55 & 0.41 & 32.91& 0.070& 0.351& 11.3 & 5.51& -- & -- & --\\
         \midrule
         Beats& LatCH-B& 89.43& 0.55& 36.77& --& --&0.138& 17.6& 5.59& {4.5} & 4.6 & 4.0 \\
         Beats& LatCH-F& 101.24& 0.7& 33.49& --& --&0.161& 17.6& 5.59& 4.1 & 4.1 & 2.9 \\
         Beats& Readout& 97.79& 0.61& 36.96& --& --&0.209& 15.7& 5.59& 4.4 &  4.7  &  2.4 \\
         Beats& End-to-end & 87.49 & 0.52 & 37.47 & --& --& 0.200 & 150.1& 30.42& {4.5} &  4.6  & 3.0  \\
         \midrule
         Beats+Intensity& LatCH-B& 87.23& 0.54& 36.60 & 4.79&   &0.141& 19.5& 5.61& 4.1  & 4.1 & 4.3 / 4.5 \\
         Beats+Intensity& LatCH-F& 92.9& 0.66& 36.27& 15.58& --&0.254& 21.4& 5.73&  3.1 & 3.9 & 4.3 / 3.4 \\
         Beats+Intensity& Readout& 103.53& 0.59& 37.26& 2.29& --&0.226&  16.1&5.62&  3.2 & 3.9 & 2.4 / 3.5 \\
         Beats+Intensity& End-to-end & 86.50 & 0.54 & 37.99 & 5.67& --& 0.200 & 240.0& 32.24& {4.4}  & 4.8 & 4.6 / 4.3 \\
         \midrule
         Pitch&  LatCH-B&  106.96&  0.51&  36.05&  --&  0.028& --& 17.7& 5.65&  3.3  &  3.9  & 4.3  \\
         Pitch&  LatCH-F&  105.57&  0.59&  35.82&  --&  0.041& --& 17.7&5.65& 3.6   &   3.7 &  3.3 \\
         Pitch&  Readout&  107.45&  0.54&  37.08&  --&  0.038& --& 15.7& 5.66&  3.5 &  3.5  &  1.3 \\
         Pitch& End-to-end & 163.24 & 0.75 & 32.75 & --& 0.030& -- & 173.1& 35.61&  3.7  &  4.5  & 3.1 \\
         \midrule
         Pitch+Intensity&  LatCH-B&  125.7&  0.49&  35.46&  3.35&  0.030& --& 19.5& 5.69&  3.1 & 4.0  & 4.7 / 3.8 \\
         Pitch+Intensity&  LatCH-F&  77.86&  0.58&  37.36&  15.27&  0.057& --& 21.3& 5.80& 3.7  &  3.9 & 1.7 / 1.5 \\
         Pitch+Intensity&  Readout&  104.45&  0.53&  38.34&  1.59&  0.047& --& 16.1& 5.69&  3.5 &  3.8 & 3.1 / 3.5 \\
         Pitch+Intensity& End-to-end & 147.64 & 0.65 & 33.95 & 1.76 & 0.033& -- & 261.1& 37.23& 3.8  & 4.1  & 3.5 / 4.3 \\
         \midrule
         Intensity&  LatCH-B&  77&  0.54&  39.23&  2.52&  --& --& 17.5& 5.56&  4.7  &  4.7  &  4.5  \\
         Intensity&  LatCH-F&  69.9&  0.55&  38.62&  15.35&  --& --& 17.5& 5.56&  4.1  &  3.9  &  2.2 \\
         Intensity&  Readout&  89.81&  0.56&  38.55&  1.38&  --& --& 15.6& 5.57&  4.7  &  4.5  &  4.7 \\
         Intensity& End-to-end & 80.76 & 0.53 & 39.62 & 2.14 & --& -- & 103.0& 26.31&  4.6  &  4.5  & 4.4  \\
         \bottomrule
    \end{tabular}
    }
    \label{tab:mainres}
\end{table*}

\subsection{Baselines}

\begin{itemize}
    \item \textit{Stable Audio Open (SAO)} \cite{evans2024open} is our audio quality baseline, as we aim to introduce control capabilities without compromising the original SAO's quality. Further, as SAO generates audio without following any control, we also set it as a lower, \emph{random} baseline for evaluating control alignment.
    \item \textit{End-to-end} is a compute-intensive baseline, similar to~\cite{Levy2023ControllableMP}, that incurs the computational burden of decoding. It is relevant for studying the trade-off between compute efficiency and audio quality (when comparing against LatCH).
    \item \textit{Readouts} \cite{luo2024readoutguidance} are similar to LatCH. We compare against them to assess there is any advantage to using intermediate layers of the diffusion model rather than the VAE latents. This being the main difference between readouts and LatCHs.
\end{itemize}

\subsection{Datasets}

We train LatCHs with the CC music from Free Music Archive that was also used to train SAO (13,874 recordings, 970h). 
Further, we evaluate the models under study with the non-vocal subset \cite{evans2024open,Evans2024LongformMG} of the Song Describer Dataset \cite{manco2023song}. 
Finally, target control signals are required for our evaluation, which can be derived from two sources: {(1)} directly extracting controls from existing data, or {(2)} creating \textit{synthetic} user-defined controls. For our quantitative evaluation we extract controls {(1)} from the Song Describer Dataset, and for our qualitative evaluation we {(2)} create \textit{synthetic} user-defined controls.

\subsection{Selective TFG, DDIM, and CFG hyperparameters}\label{sec:hyps}

Given the challenge of tuning TFG hyperparameters, we present our selected values alongside the intuitions that informed their selection.

\begin{itemize}
    \item \textit{LatCH:} $\rho$ = $\mu$ = 0.03, $\gamma$ = 0.3, and apply selective TFG only to the first 20\% sampling steps. 
    If $\rho$, $\mu$, or the \% of selected TFG steps is too high, it exists the possibility to drift off-manifold. Having small $\rho$, $\mu$, and applying TFG for a few sampling steps guarantees that the output quality is not compromised (while following the control). 
    For using multiple controls, we weight the \textit{intensity} loss by 0.0005.
    
    \item \textit{End-to-end:} $\rho$ = $\mu$ = 0.03, $\gamma$ = 1.5, and we apply selective TFG only to the first 20\% sampling steps. We found that it was key here to increase $\gamma$, as it was following the control at the expense of compromising audio quality. In this case, decreasing the guidance by attenuating $\rho$, $\mu$, or decreasing the \% of selected steps degraded control following. For this reason, we focused on increasing $\gamma$  to reduce the accuracy of the guidance. We weight the \textit{intensity} loss by 0.001.
    
    \item \textit{Readouts:} $\rho$ = 0.1. %
    Readouts use intermediate layers $\bm{h}_t$ of a latent diffusion model, \textit{i.e.}, $\tilde{\bm{e}}_{t} = \mathbf{r}(\bm{h}_t(\bm{z}_t))$. As such, the mean guidance term $\nabla_{\mathbf{z}_{0\mid t}}\mathcal{\delta}(\tilde{\bm{e}}_{0\mid t}, \bm{e})$ is zero, as $\mathbf{r}(\bm{h}_t(\bm{z}_t))$ is function of ${\bm{z}}_{t}$ (not ${\bm{z}}_{0\mid t}$). 
    Thus, readouts are not suitable for mean guidance, iteration on  mean guidance, or for convolving $\mathbf{z}_{0\mid t}$ with gaussian noise, and are only usable with variance guidance.
    We weight the \textit{intensity} loss scale by 0.005.
    
\end{itemize}

\noindent We set $N_{\mathrm{iter}} = 4$ and $N_{\mathrm{recur}} = 1$ throughout, as in \cite{ye2024tfg}. Empirically we found that our procedure is more stable using the variance-preserving form that SAO was trained in, rather than the variance-\emph{exploding} form from its public code, and thus use a uniform noise schedule, 100-step stochastic DDIM sampler
and CFG scale of 7.

We do not report hyperparameter ablations, as large hyperparameter changes were ineffective and small variations negligible. %

\subsection{Training LatCHs} 
\label{sec:traininglatchs}

Assuming that $\mathbf{C}$ and $\mathcal{E}$ outputs have the same time resolution, LatCHs are bidirectional transformers, operating on VAE latents with RoPE \cite{su2024roformer}, followed by a projection layer to the output dimension of the features. Hence, we downsample/upsample the extracted features to the same temporal resolution as the latents, as one does not need controls to vary at much faster than 21.5Hz \cite{evans2024open}.
LatCHs are of $\approx$ 7M parameters ($< 1\%$ the parameters of the base generative model \cite{evans2024open}) and train in $\approx$ 4 hours on a single H100 GPU, offering a lower-resource control method than conditional training.

Some feature extractors, \textit{e.g.}, CREPE \cite{kim2018crepe} or All-in-One \cite{kim2023all}, are neural models. We trained LatCHs to match their logits, rather than using hard-labeled targets. Further, extractors can be sparse, \textit{e.g.}, CREPE, which has 160 pitch bins, where only 1 bin is active at a time. In such cases, we use a sparsity-aware loss: given a threshold to denote non-activating features, we average the below and above threshold losses together to upweight the importance of the sparse feature. Thus, we have the following training recipes: (1)~\textit{intensity LatCHs} that train with MSE against RMS, (2)~\textit{pitch LatCHs} that train with sparse-weighted BCE (threshold=0.2) against CREPE logits, and (3)~\textit{beats LatCHs} that train with BCE against All-in-One logits.

\subsection{Quantitative Evaluation}

Our evaluation is based on established \cite{stableaudio} quality metrics %
that include FD\textsubscript{openl3}~\cite{cramer2019look}, KL\textsubscript{passt}~\cite{koutini2021efficient}, and CLAP~\cite{wu2023large}, which measures text-audio similarity.
We also evaluate how close the generated audio is able to match the controls based on their distance function $\mathcal{\delta}(\cdot)$: MSE for \textit{intensity} and BCE for \textit{pitch} and \textit{beats}. Finally, we also evaluate runtime and VRAM usage on an H100 GPU.

\subsection{Qualitative evaluation}

We evaluate our method qualitatively with a listening test. Participants ($n=15$) were asked to rate examples on audio quality, prompt adherence, and control alignment.
We report mean opinion scores (MOS) on a 5-point scale from 1 (bad) to 5 (excellent).

\section{Results and discussion}
\label{sec:results}

Next, we discuss our qualitative and quantitative results (in Table 1):

\begin{itemize}
    \item {LatCH-B} achieves the best performance in audio quality, prompt adherence, control alignment, and efficiency.
    \item End-to-end also performs well, but at a high compute cost.
    \item Readouts tend to be worse in our study, possibly because it lacks the ``mean guidance''-related terms in TFG (Sec. \ref{sec:traininglatchs}).
    \item MOS results are generally {good} for \textit{beats}, \textit{intensity}, end-to-end \& LatCH-B, and quality metrics are comparable to SAO.
    \item Multiple guidance controls also works (\textit{e.g.}, \textit{beats}+\textit{intensity}).
\end{itemize}

\noindent We find that the studied methods are more reliable with gradual or low-frequency controls, such as \textit{intensity} or \textit{beats}. 
Controls with greater variability, such as \textit{pitch}, which involves rapid note changes, pose challenges, and show worse performance across quality metrics (see Table 1).
Also note that \textit{intensity} and \textit{beats} consist of 1D outputs and \textit{pitch} outputs 160 pitch classes.
We also explored PANNs (527 tags) and Chroma (12 pitches) feature extractors with mixed results. Based on those observations, we hypothesize that 1D outputs, rather than sparse high-dimensional ones, can be more suited for guidance.

\section{Conclusions}

We present a low-resource, inference-time control framework for latent audio diffusion models by combining selective TFG with lightweight, trainable latent-control heads (LatCHs). Our method enables control without re-training the base generative model or incurring the high computational cost of end-to-end guidance. Because LatCHs operate directly in latent space, is not required to run backpropagation through the decoder as end-to-end guidance does. Further, we introduce selective TFG, which strategically applies guidance at few, selected diffusion steps, improving efficiency while preserving generation quality. Our approach effectively balances control precision, audio fidelity, and runtime performance, enabling long-form, steerable audio synthesis (up to 47.55s). Our method can also use multiple control signals and features at the same time.

\bibliographystyle{IEEEbib}
\bibliography{refs25}

\end{document}